\documentclass[reprint,prd,twocolumn,superscriptaddress]{revtex4-1}

\usepackage{color}
\usepackage{amsmath}
\usepackage{graphicx}

\allowdisplaybreaks

\newcommand{\evenspacing}{\vphantom{\bigg[}}
\begin{document}

\mbox{}\hfill{\small CP3-Origins-2019-47 DNRF90, TTP19-047, P3H-19-053}

\title{Gauge coupling beta functions to four-loop order in the
  Standard Model}
\author{Joshua Davies}
\email{joshua.davies@kit.edu}
\author{Florian Herren}
\email{florian.herren@kit.edu}
\affiliation{Institut f\"ur Theoretische Teilchenphysik,
  Karlsruhe Institute of Technology (KIT), 76128 Karlsruhe, Germany}

\author{Colin Poole}
\email{cpoole.ac@gmail.com}
\affiliation{Unlisted, formerly CP$^3$-Origins, University of Southern Denmark,
  Campusvej 55, DK-5230 Odense M, Denmark}

\author{Matthias Steinhauser}
\email{matthias.steinhauser@kit.edu}
\affiliation{Institut f\"ur Theoretische Teilchenphysik,
  Karlsruhe Institute of Technology (KIT), 76128 Karlsruhe, Germany}

\author{Anders Eller Thomsen}
\email{aethomsen@cp3.sdu.dk}
\affiliation{CP$^3$-Origins, University of Southern Denmark,
  Campusvej 55, DK-5230 Odense M, Denmark}

\begin{abstract}
  We compute the beta functions of the three Standard Model gauge 
  couplings to four-loop order in the modified minimal subtraction scheme.
  At this order a proper definition of $\gamma_5$ in $D=4-2\epsilon$
  space-time dimensions is required; however, in our
  calculation we determine the $\gamma_5$-dependent terms by exploiting relations with
  beta function coefficients at lower loop orders.
\end{abstract}
\maketitle


{\bf Introduction.}
Beta functions are fundamental quantities of quantum field theories.
They are important ingredients of the renormalization group equations
and determine the energy dependence of the couplings.  The perturbative
coefficients that are currently available enter into a variety of
applications, among which is the running of the Standard Model (SM)
couplings from the electroweak scale to the scale where the coupling
of the quartic terms in the scalar potential turns negative and
the vacuum becomes unstable~\cite{Bezrukov:2012sa,Degrassi:2012ry,Alekhin:2012py}.
A precise running of the coupling constants is also
needed in the context of the prediction of Higgs boson masses
within the Minimal Supersymmetric extension of the SM (MSSM).
In the approach discussed, e.g., in Ref.~\cite{Bagnaschi:2019esc},
all SM quantities are evolved to the supersymmetric scale,
which is usually of the order of a few TeV, where the
matching between the SM and the MSSM is performed.

The gauge structure of the SM of particle physics is given by
$\mathrm{SU}(3)\times \mathrm{SU}(2) \times \mathrm{U}(1)$, and thus there are
three gauge couplings.  In this letter we compute their beta functions to
four-loop accuracy, with the only approximation that the Yukawa couplings of
the first and second generations are set to zero.  For our calculation we
adopt the widely-used $\overline{\rm MS}$ scheme. Furthermore,
since the beta functions are mass-independent, we can work in the unbroken
phase of the SM in which all particles are massless.

Within the SM a number of correction terms to the various beta functions are
available. The discovery of asymptotic freedom in non-abelian gauge
theories~\cite{Gross:1973id,Politzer:1973fx} prompted the computation of
two-loop corrections within the strong sector of the SM, which became available
shortly afterwards~\cite{Jones:1974mm,Caswell:1974gg,Tarasov:1976ef,Egorian:1978zx}.
Three- and four-loop corrections have been computed
in~\cite{Tarasov:1980au,Larin:1993tp}
and~\cite{vanRitbergen:1997va,Czakon:2004bu} respectively, and recently
even the five-loop term became
available~\cite{Baikov:2016tgj,Herzog:2017ohr,Luthe:2017ttg}.

Two-loop corrections to the beta functions of all couplings of the SM can be
found in
Refs.~\cite{Jones:1981we,Machacek:1983tz,Machacek:1983fi,Machacek:1984zw},
and the three-loop corrections to all gauge coupling beta functions have been
computed in~\cite{Mihaila:2012fm,Mihaila:2012pz,Bednyakov:2012rb}.  The
three-loop Yukawa coupling beta functions have been considered
in~\cite{Chetyrkin:2012rz,Bednyakov:2012en,Bednyakov:2014pia,Herren:2017uxn}
and the scalar self coupling beta functions
in~\cite{Chetyrkin:2013wya,Bednyakov:2013eba,Bednyakov:2013cpa}.  At four-loop
order partial results are available;
in~\cite{Martin:2015eia,Chetyrkin:2016ruf} the scalar self coupling beta
function and in~\cite{Bednyakov:2015ooa,Zoller:2015tha} the top quark Yukawa
contributions to the QCD beta function have been computed.

In the approximation that the Yukawa couplings of the first and second
generation fermions are neglected, the SM has seven couplings.
Their beta functions are defined as
\begin{align}
  \mu^2\frac{\mathrm{d}}{\mathrm{d}\mu^2}\frac{\alpha_i}{\pi} =
  \beta_i\left(\left\{\alpha_j\right\},\epsilon\right)
\end{align}
with $i=1,\ldots,7$, where $d = 4-2\epsilon$ is the space-time dimension,
$\mu$ is the renormalization scale
and $\{\alpha_j\}$ denotes dependence on all seven couplings.
$\alpha_1$, $\alpha_2$ and $\alpha_3$ are the three gauge couplings,
which we define using a $SU(5)$-like normalization
\begin{align}
  \alpha_1 = \frac{5}{3}\frac{\alpha_{\mathrm{QED}}}{\cos^2\theta_W}~,\quad
  \alpha_2 = \frac{\alpha_{\mathrm{QED}}}{\sin^2\theta_W}~,\quad
  \alpha_3 = \alpha_s~,
\end{align}
where $\alpha_{\mathrm{QED}}$ is the fine structure constant,
$\theta_W$ is the weak mixing angle and $\alpha_s$ is the strong
coupling constant.
In order to fix the Yukawa couplings, we provide the
corresponding part of the Lagrange density,
\begin{align}
  \mathcal{L} \supset
  y_t \overline{Q}_L\left(\mathrm{i}\tau_2\Phi^*\right)t_R
  + y_b \overline{Q}_L\Phi b_R
  + y_\tau \overline{L}_L\Phi \tau_R
  + \mathrm{h.c.}
\end{align}
where $\tau_2$ is the second Pauli matrix, $Q_L$ and $L_L$ are the 
3rd generation left-handed quark and lepton doublets, $\Phi$ the Higgs doublet and
$t_R$,~$b_R$,~$\tau_R$ the right-handed top, bottom and $\tau$
fields. We use the coupling factors $y_i$ to define 
the third-generation Yukawa couplings as
\begin{align}
  \alpha_4 = \frac{y_t^2}{4\pi}~,\quad
  \alpha_5 = \frac{y_b^2}{4\pi}~,\quad
  \alpha_6 = \frac{y_\tau^2}{4\pi}~.
\end{align}
Finally, we provide the quartic term of the scalar potential,
which fixes $\alpha_7$:
\begin{align}
  \mathcal{L} \supset
  -\left(4\pi\alpha_7\right)\left(\Phi^\dagger\Phi\right)^2~.
\end{align}

The beta functions are obtained from the renormalization constants
using the formula (see, e.g., \cite{Mihaila:2012fm,Mihaila:2012pz})
\begin{align}
  \beta_i = - \Bigg[\epsilon\frac{\alpha_i}{\pi} +
    \frac{\alpha_i}{Z_{\alpha_i}}\sum_{j=1,\:j\neq i}^7\frac{\partial
      Z_{\alpha_i}}{\partial \alpha_j}\beta_j\Bigg] \! \Bigg(1 +
  \frac{\alpha_i}{Z_{\alpha_i}}\frac{\partial Z_{\alpha_i}}{\partial
    \alpha_i}\Bigg)^{-1}\!\!,
  \label{eq::beta_Z}
\end{align}
where the renormalization constants are obtained from the
relations between the bare and renormalized couplings,
\begin{align}
  \alpha_i^0 =
  \mu^{2\epsilon}Z_{\alpha_i}\left(\left\{\alpha_j\right\},\epsilon\right)
  \alpha_i~.
\end{align}
Note that the Yukawa and self
couplings enter the gauge coupling renormalization constants for the first
time at two- and three-loop order, respectively. Thus, from
Eq.~(\ref{eq::beta_Z}) one learns that the four-loop gauge coupling beta
functions require the knowledge of the two-loop Yukawa coupling beta functions
and one-loop beta function for $\alpha_7$.

\smallskip

\begin{figure}[t]
  \begin{center}
    \includegraphics[width=.35\textwidth]{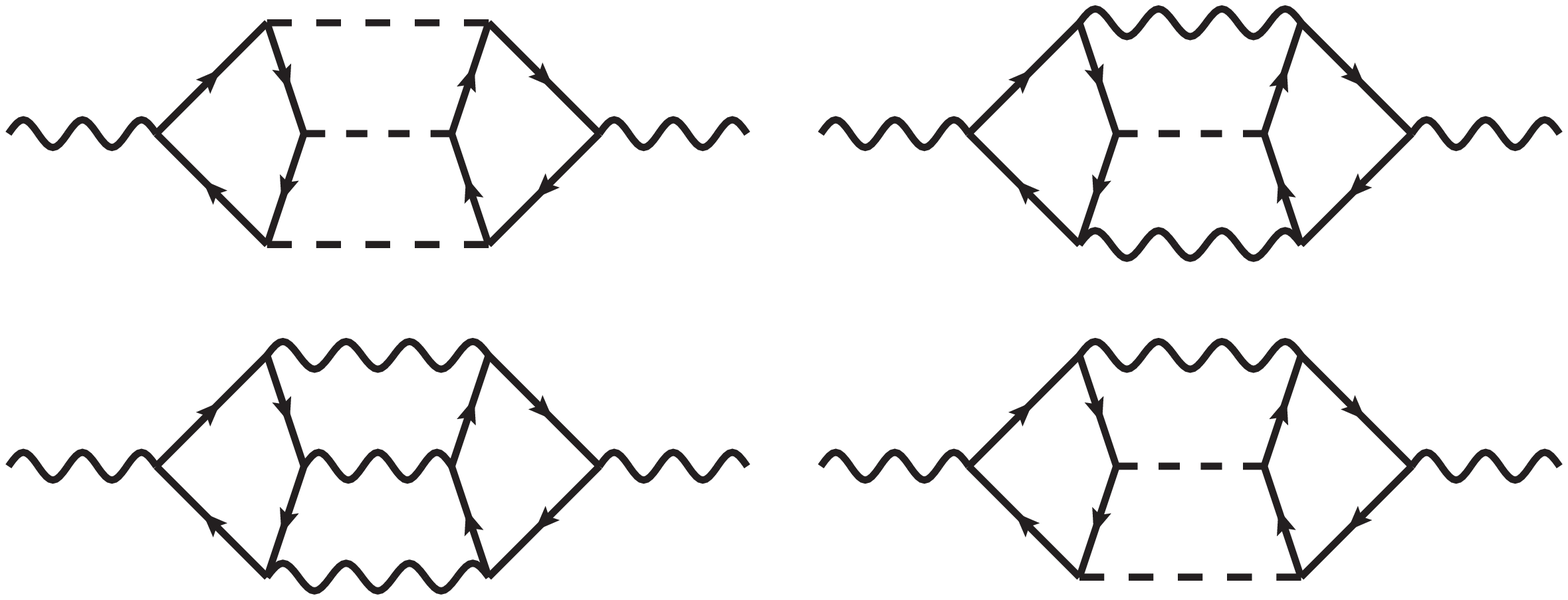}
    \caption{\label{fig::g54l} Representative four-loop diagrams for classes
      that might involve non-trivial $\gamma_5$ contributions.
      Wavy, dashed and straight lines represent gauge bosons, scalar bosons
      and fermions respectively.}
  \end{center}
\end{figure}

{\bf Weyl consistency conditions.}
As we will discuss below, the computation of the renormalization
constants can be reduced to the evaluation of massless four-loop
two-point functions. Although methods for this have been available
for a few years, to date the four-loop corrections to the beta functions
in the electroweak sector have not been computed. The main
reason for this is connected to traces containing an odd number
of $\gamma_5$ matrices: whereas at three-loop order a semi-naive
treatment is possible, a proper treatment is (in principle) required at
four loops. 
The classes of diagrams that might require such a treatment
need to have at least two (open or closed) fermion lines
with sufficiently many vertices.
In our case, only the diagram classes shown in Fig.~\ref{fig::g54l}
satisfy this criterion.
For massless fermions the
diagrams in the top row are zero, since all traces involve an odd
number of gamma matrices. Furthermore, in the left diagram in the
second row the dangerous contributions cancel due to anomaly
cancellations within the SM.
This leaves only the class of diagrams
with two fermion loops that are connected by one vector and two
scalar bosons.  In Refs.~\cite{Bednyakov:2015ooa,Zoller:2015tha} such
diagrams have been considered for the case where the gauge boson is a
gluon.  In order to treat the problematic traces, the cyclicity of the
traces was abandoned, and different results were obtained depending on
what starting point was used to write down the traces.

In the literature one finds various prescriptions for the treatment of
$\gamma_5$ in $D$ dimensions, see, e.g.,
Refs.~\cite{tHooft:1972tcz,Korner:1991sx,Larin:1993tq,Jegerlehner:2000dz,Zerf:2019ynn}.
Many of these have been successfully applied in various calculations
either in pure QCD or at lower loop order.
In our opinion there is no
practical prescription that can be applied at fourth order in perturbation
theory. However, very recently in Ref.~\cite{Poole:2019txl,Poole:2019kcm}
Weyl consistency conditions~\cite{Osborn:1989td,Jack:1990eb,Osborn:1991gm,Jack:2013sha} have been used
in order to establish, with the help of ``Osborn's equation'',
relations between coefficients of the general four-loop gauge, three-loop Yukawa and
two-loop scalar beta function. It was realized
in~\cite{Poole:2019txl,Poole:2019kcm} that these relations fix all non-trivial
$\gamma_5$ contributions to the four-loop gauge coupling beta function in
terms of known coefficients of the three-loop Yukawa beta function.  In
particular, the results of~\cite{Poole:2019txl,Poole:2019kcm} could resolve
the ambiguity of the four-loop top Yukawa contribution to the beta function of
the strong coupling constant, identified in~\cite{Bednyakov:2015ooa,Zoller:2015tha}.

This observation fixes the outline for our computation; we decompose
the beta functions into colour structures of the three gauge
groups. We then perform an explicit computation of those parts of the
renormalization constants that do not involve traces with an odd
number of $\gamma_5$ matrices and fix the remaining parts using the
results obtained in~\cite{Poole:2019txl,Poole:2019kcm}.  In
addition, in Refs.~\cite{Poole:2019txl,Poole:2019kcm} many further
relations have been established that demonstrate the consistency between
predictions derived from Osborn's equation and explicit computations. 

\smallskip

\begin{figure}[t]
  \begin{center}
    \includegraphics[width=.45\textwidth]{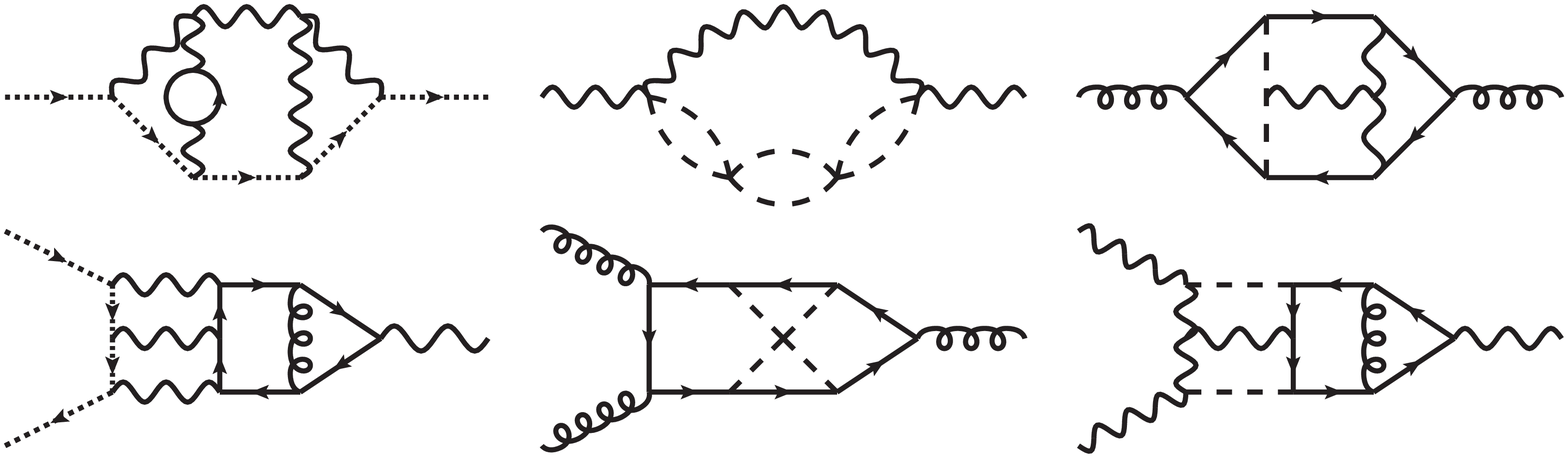}
    \caption{\label{fig::diags}Sample Feynman diagrams contributing to 
      the Green's functions that have been used for our calculation of the 
      gauge coupling renormalization constants.
      Solid, dashed, dotted, curly and wavy lines denote fermions,
      scalar bosons, ghosts, gluons and electroweak gauge bosons
      respectively.} 
  \end{center}
\end{figure}

{\bf Calculation.}
For the computation of the gauge coupling renormalization constants,
one can, in principle, use any vertex that contains the respective
coupling at tree level. The renormalization constant is then
obtained by
\begin{eqnarray}
  Z_{g_i} &=& \frac{ Z_{\rm vert} }{\Pi_{k} \sqrt{Z_{k,{\rm wf}}} }
  \,,
\end{eqnarray}
where $Z_{\rm vert}$ stands for the renormalization constant of the vertex and
$Z_{k,{\rm wf}}$ for the wave function renormalization constants ($k$ runs
over all external particles).  For the $\mathrm{SU}(2)$ and $\mathrm{SU}(3)$ gauge groups, it is
advantageous to choose the ghost--gauge-boson vertices since one has to deal with
fewer diagrams, amounting to ${\cal O}(350,000)$ for $\mathrm{SU}(2)$ and
${\cal O}(200,000)$ for $\mathrm{SU}(3)$ at four-loop order.  For the $\mathrm{U}(1)$ gauge
group, it is sufficient to consider the gauge boson propagator renormalization
constant for which ${\cal O}(200,000)$ four-loop diagrams have to be computed.
Sample Feynman diagrams for the various Green's functions we
consider are shown in Fig.~\ref{fig::diags}.

Our calculation is based on a well-tested setup, which uses {\tt
qgraf}~\cite{Nogueira:1991ex} for the generation of the amplitudes and {\tt
q2e} and {\tt exp}~\cite{Harlander:1997zb,Seidensticker:1999bb,q2eexp} for
the mapping to integral families and generation of {\tt
  FORM}~\cite{Ruijl:2017dtg} code. We use {\tt
  color}~\cite{vanRitbergen:1998pn} for the computation of the $\mathrm{SU}(2)$ and $\mathrm{SU}(3)$
colour factors. Before the computation we combine diagrams with the
same colour structure and integral family to form so-called ``superdiagrams'', which
guarantees possible cancellations at earlier stages of the calculation.
This reduces the computational effort required.

In the unbroken phase of the SM, all particles are massless, and thus all
two-point Green's functions lead to massless propagator-type integrals up to
four loops. Furthermore, one may set one of the external momenta of
the three-point Green's functions to zero. This is possible since the diagrams are
logarithmically divergent, and thus the ultra-violet divergences, which must be
computed to obtain the renormalization constants in the $\overline{\rm MS}$
scheme, are independent of kinematic quantities.
Consequently, one only needs to compute massless
propagator-type integrals up to four loops;  this also holds for vertex corrections.
For this task we use the program \texttt{FORCER}~\cite{Ruijl:2017cxj}.

In our calculation we use an anti-commuting $\gamma_5$, with $\gamma_5^2 = 1$, and
set traces with an odd number of $\gamma_5$ occurrences to zero. These
contributions to the beta functions (and thus to the gauge coupling
renormalization constants) are reconstructed using the approach
described in the previous section. Note that the ghost--gauge-boson vertices do not
suffer from ambiguities related to $\gamma_5$. This
allows us to reconstruct the non-trivial $\gamma_5$ contributions 
to the renormalization constants of the gauge boson wave-functions.

We retain full dependence on all three gauge parameters during the
calculation.  Whereas the renormalization constants for the vertices and wave
functions still depend on the gauge parameters, the dependence drops out in the
renormalization constants of the gauge couplings. This serves as a welcome
check of our calculation.

As a further strong check we use the triple gauge boson vertices for the
$\mathrm{SU}(2)$ and $\mathrm{SU}(3)$ gauge bosons to re-compute the gauge coupling
renormalization constants, and find agreement.
Furthermore, we verify by explicit calculation that $\mathrm{U}(1)$ loop corrections
to the triple gauge boson vertex vanish after all bare
Feynman diagrams are added. The calculation of these Green's functions are
significantly more costly than our default choice; for this reason we fix the
gauge parameters to the Feynman gauge for these Green's functions only.

We have performed several cross-checks of our four-loop expressions with results
available in the literature.  The pure gauge-fermion parts of $\beta_{2}$ and
$\beta_{3}$ agree with the findings for a general Yang-Mills theory with
fermions in the fundamental
representation~\cite{vanRitbergen:1997va,Czakon:2004bu}.  Furthermore, the
contributions to $\beta_3$ involving only the strong gauge coupling, the top
quark Yukawa coupling and the quartic scalar coupling agree
with~\cite{Bednyakov:2015ooa,Zoller:2015tha}.

Finally, the Weyl consistency conditions from
Refs.~\cite{Poole:2019txl,Poole:2019kcm} represent powerful cross-checks on
various coefficients in the beta functions, via their relation to the general
result. The parametrization of the general four-loop gauge beta function
(valid for all renormalizable four-dimensional quantum field theories) has 202 coefficients. As
mentioned above, the four calculated in \cite{Poole:2019txl} determine all
contributions from traces over an odd number of $ \gamma_5 $ matrices, and are
used directly in the computation of the beta functions. While we do not yet
have a complete determination of the other 198, matching the general result to
our SM calculation does uniquely fix 80; comparison with the full set of 261
consistency conditions in \cite{Poole:2019kcm} (which also involve
coefficients of the general three-loop Yukawa beta function) verifies these 80,
and fixes another 28.
Crucially, we find that these 108 coefficients (and by extension our
four-loop computation) are indeed consistent with all Weyl
consistency conditions, providing highly non-trivial corroboration.

\smallskip

{\bf Results.}
Our final results for the gauge coupling beta functions contain the full
dependence on the gauge and Higgs self couplings and the third generation
Yukawa couplings. The analytic results are available in computer-readable
from~\cite{progdata}.  Due to space restrictions we reproduce below the
results for vanishing bottom and tau Yukawa couplings, $\alpha_5=\alpha_6=0$,
and for numerical values of the $\mathrm{SU}(2)$ and $\mathrm{SU}(3)$ Casimir invariants.
They are given by
\begin{widetext}
\begin{align}
&\beta_1 =
\frac{\alpha^2_1}{\left(4\pi\right)^2}\left(\frac{82}{5}\right) + \frac{\alpha^2_1}{\left(4\pi\right)^3}\left(\frac{398 \alpha_1}{25} + \frac{54 \alpha_2}{5} + \frac{176 \alpha_3}{5} - \frac{34 \alpha_4}{5}\right) + 
\frac{\alpha^2_1}{\left(4\pi\right)^4}\bigg(-\frac{388613 \alpha_1^2}{6000} + \frac{123 \alpha_1 \alpha_2}{40} - \frac{548 \alpha_1 \alpha_3}{75} 
\nonumber\\&\evenspacing
+ \frac{789 \alpha_2^2}{16} 
- \frac{12 \alpha_2 \alpha_3}{5} + \frac{1188 \alpha_3^2}{5} - \frac{2827 \alpha_1 \alpha_4}{200} - \frac{471 \alpha_2 \alpha_4}{8} - \frac{116 \alpha_3 \alpha_4}{5} + \frac{189 \alpha_4^2}{4} + \frac{54 \alpha_1 \alpha_7}{25}
+ \frac{18 \alpha_2 \alpha_7}{5} - \frac{36 \alpha_7^2}{5}\bigg)
\nonumber\\&\evenspacing
+ \frac{\alpha^2_1}{\left(4\pi\right)^5}\bigg[-\alpha_1^3
  \left(\frac{143035709}{1080000} + \frac{1638851 \zeta_3}{5625}\right) -
  \alpha_1^2  \alpha_2\left(\frac{3819731}{24000} - \frac{16529 \zeta_3}{125}\right) 
 - \alpha_1^2  \alpha_3\left(\frac{3629273}{6750} - \frac{720304 \zeta_3}{1125}\right)
\nonumber\\&\evenspacing
+ \alpha_1  \alpha_2^2\left(\frac{572059}{14400} - \frac{6751 \zeta_3}{75}\right) - \frac{69  \alpha_1  \alpha_2  \alpha_3}{25} +  \alpha_1\alpha_3^2\left(\frac{333556}{675} - \frac{274624 \zeta_3}{225}\right)
- \alpha_2^3\left(\frac{117923}{2880} + \frac{3109 \zeta_3}{5}\right)
\nonumber\\&\evenspacing
 - \alpha_2^2  \alpha_3\left(\frac{41971}{90} - \frac{7472 \zeta_3}{15}\right)-  \alpha_2  \alpha_3^2 \left(\frac{1748}{3} - \frac{2944 \zeta_3}{5}\right)  +  \alpha_3^3 \left(\frac{6116}{15} - \frac{18560 \zeta_3}{9}\right) + 
   \alpha_1^2 \alpha_4\left(\frac{8978897}{72000} + \frac{2598 \zeta_3}{125}\right) 
\nonumber\\&\evenspacing
- \alpha_1 \alpha_2 \alpha_4 \left(\frac{42841}{800} + \frac{1122 \zeta_3}{25}\right)- \alpha_1 \alpha_3 \alpha_4 \left(\frac{2012}{75} - \frac{408 \zeta_3}{25}\right) 
- \alpha_2^2 \alpha_4\left(\frac{439841}{960} - \frac{616 \zeta_3}{5}\right) + \alpha_2 \alpha_3 \alpha_4 \left(\frac{1468}{5} - \frac{1896 \zeta_3}{5}\right) 
\nonumber\\&\evenspacing
- \alpha_3^2 \alpha_4 \left(\frac{11462}{45} - \frac{3184 \zeta_3}{5}\right)  + \alpha_1 \alpha_4^2 \left(\frac{29059}{160} - \frac{357 \zeta_3}{25}\right) + \alpha_2 \alpha_4^2 \left(\frac{71463}{160} + \frac{639 \zeta_3}{5}\right) + \alpha_3 \alpha_4^2 \left(\frac{1429}{5} - 240 \zeta_3\right)
\nonumber\\&\evenspacing
 - \alpha_4^3 \left(\frac{13653}{40} + \frac{102 \zeta_3}{5}\right) + \frac{3627 \alpha_1^2 \alpha_7}{500}
+ \frac{1917 \alpha_1 \alpha_2 \alpha_7}{50} + \frac{889 \alpha_2^2 \alpha_7}{20} - \frac{1926 \alpha_1 \alpha_4 \alpha_7}{25} - \frac{162 \alpha_2 \alpha_4 \alpha_7}{5} - \frac{474 \alpha_4^2 \alpha_7}{5}
\nonumber\\&\evenspacing
-\frac{1269 \alpha_1 \alpha_7^2}{25} - \frac{981 \alpha_2 \alpha_7^2}{5}+ \frac{1188 \alpha_4 \alpha_7^2}{5} + \frac{624 \alpha_7^3}{5}\bigg]~,
\\
\nonumber\\
&\beta_2 =
\frac{\alpha^2_2}{\left(4\pi\right)^2}\left(-\frac{38}{3}\right) + \frac{\alpha^2_2}{\left(4\pi\right)^3}\left(\frac{18 \alpha_1}{5} + \frac{70 \alpha_2}{3} + 48 \alpha_3 - 6 \alpha_4\right) + 
\frac{\alpha^2_2}{\left(4\pi\right)^4}\bigg(-\frac{5597 \alpha_1^2}{400} + \frac{873 \alpha_1 \alpha_2}{40} - \frac{4 \alpha_1 \alpha_3}{5}
\nonumber\\&\evenspacing
 + \frac{324953 \alpha_2^2}{432}
+ 156 \alpha_2 \alpha_3 + 324 \alpha_3^2 - \frac{593 \alpha_1 \alpha_4}{40} - \frac{729 \alpha_2 \alpha_4}{8} - 28 \alpha_3 \alpha_4 + \frac{147 \alpha_4^2}{4} + \frac{6 \alpha_1 \alpha_7}{5} + 6 \alpha_2 \alpha_7 - 12 \alpha_7^2\bigg)
\nonumber\\&\evenspacing
+ \frac{\alpha^2_2}{\left(4\pi\right)^5}\bigg[-\alpha_1^3 \left(\frac{6418229}{72000} - \frac{21173 \zeta_3}{375}\right) -  \alpha_1^2  \alpha_2\left(\frac{787709}{4800} - \frac{659 \zeta_3}{25}\right) 
 - \alpha_1^2  \alpha_3\left(\frac{52297}{450} - \frac{2032 \zeta_3}{15}\right) + \frac{161  \alpha_1  \alpha_2  \alpha_3}{5}
\nonumber\\&\evenspacing
- \alpha_1  \alpha_2^2\left(\frac{375767}{2880} - \frac{4631 \zeta_3}{15}\right) -  \alpha_1\alpha_3^2\left(\frac{1748}{9} - \frac{2944 \zeta_3}{15}\right)
+ \alpha_2^3\left(\frac{124660945}{15552} - \frac{78803 \zeta_3}{9}\right) - \alpha_2^2  \alpha_3\left(\frac{72881}{18} - \frac{16432 \zeta_3}{3}\right)
\nonumber\\&\evenspacing
+  \alpha_2  \alpha_3^2 \left(\frac{10348}{3} - 2560 \zeta_3\right)  +  \alpha_3^3 \left(\frac{1028}{3} - \frac{7040 \zeta_3}{3}\right) + 
   \alpha_1^2 \alpha_4\left(\frac{465089}{4800} - \frac{498 \zeta_3}{25}\right) - \alpha_1 \alpha_2 \alpha_4 \left(\frac{102497}{480} + 28 \zeta_3\right)
\nonumber\\&\evenspacing
+ \alpha_1 \alpha_3 \alpha_4 \left(\frac{796}{15} - \frac{376 \zeta_3}{5}\right) 
- \alpha_2^2 \alpha_4\left(\frac{500665}{576} - \frac{478 \zeta_3}{3}\right) - \alpha_2 \alpha_3 \alpha_4 \left(\frac{1444}{3} + 56 \zeta_3\right) - \alpha_3^2 \alpha_4 \left(\frac{614}{3} - 336 \zeta_3\right)
\nonumber\\&\evenspacing
  + \alpha_1 \alpha_4^2 \left(\frac{3161}{32} + \frac{153 \zeta_3}{5}\right) + \alpha_2 \alpha_4^2 \left(\frac{30213}{32} - 63 \zeta_3\right) + \alpha_3 \alpha_4^2 \left(239 - 144 \zeta_3\right)
 - \alpha_4^3 \left(\frac{2143}{8} + 18 \zeta_3\right) + \frac{457 \alpha_1^2 \alpha_7}{100}
\nonumber\\&\evenspacing
+ \frac{69 \alpha_1 \alpha_2 \alpha_7}{2} + \frac{2905 \alpha_2^2 \alpha_7}{12} - \frac{54 \alpha_1 \alpha_4 \alpha_7}{5} - 150 \alpha_2 \alpha_4 \alpha_7 - 78 \alpha_4^2 \alpha_7
 - \frac{327 \alpha_1 \alpha_7^2}{5} - 363 \alpha_2 \alpha_7^2 + 300 \alpha_4 \alpha_7^2 + 208 \alpha_7^3
\bigg]\,,
\\
\nonumber\\
&\beta_3 =
\frac{\alpha^2_3}{\left(4\pi\right)^2}\left(-28\right) + \frac{\alpha^2_3}{\left(4\pi\right)^3}\left(\frac{22\alpha_1}{5} + 18\alpha_2 - 104\alpha_3 - 8\alpha_4\right) + 
\frac{\alpha^2_3}{\left(4\pi\right)^4}\bigg(-\frac{523 \alpha_1^2}{30} - \frac{3 \alpha_1 \alpha_2}{10} + \frac{109 \alpha_2^2}{2} + \frac{308 \alpha_1 \alpha_3}{15}
\nonumber\\&\evenspacing
+ 84 \alpha_2 \alpha_3 + 130 \alpha_3^2 - \frac{101 \alpha_1 \alpha_4}{10} - \frac{93 \alpha_2 \alpha_4}{2} - 160 \alpha_3 \alpha_4 + 60 \alpha_4^2\bigg)
\nonumber\\&\evenspacing
+ \frac{\alpha^2_3}{\left(4\pi\right)^5}\bigg[-\alpha_1^3 \left(\frac{6085099}{54000} - \frac{17473 \zeta_3}{225}\right) - \alpha_1^2  \alpha_2\left(\frac{46951}{1200} - \frac{973 \zeta_3}{25}\right) 
-  \alpha_1^2  \alpha_3\left(\frac{35542}{135} - \frac{902 \zeta_3}{9}\right) + \frac{69  \alpha_1  \alpha_2  \alpha_3}{5}
\nonumber\\&\evenspacing
- \alpha_1  \alpha_2^2\left(\frac{37597}{720} - \frac{691 \zeta_3}{15}\right) - \alpha_1\alpha_3^2\left(\frac{57739}{135} - \frac{32476 \zeta_3}{45}\right)
 - \alpha_2^3\left(\frac{176815}{432} + 935 \zeta_3\right) +  \alpha_2^2  \alpha_3\left(\frac{3812}{9} - \frac{950 \zeta_3}{3}\right)
\nonumber\\&\evenspacing
- \alpha_2  \alpha_3^2 \left(\frac{5969}{3} - 3476 \zeta_3\right)  +  \alpha_3^3 \left(\frac{127118}{9} - \frac{179792 \zeta_3}{9}\right) + 
   \alpha_1^2 \alpha_4\left(\frac{362287}{3600} - \frac{19 \zeta_3}{25}\right) + \alpha_1 \alpha_2 \alpha_4 \left(\frac{77}{40} - 54 \zeta_3\right)
\nonumber\\&\evenspacing
- \alpha_1 \alpha_3 \alpha_4 \left(\frac{1283}{15} + \frac{32 \zeta_3}{5}\right) 
- \alpha_2^2 \alpha_4\left(\frac{12887}{48} - 117 \zeta_3\right) - \alpha_2 \alpha_3 \alpha_4 \left(473 + 288 \zeta_3\right) - \alpha_3^2 \alpha_4 \left(\frac{26836}{9} - 1088 \zeta_3\right)
\nonumber\\&\evenspacing
 + \alpha_1 \alpha_4^2 \left(\frac{3641}{40} + \frac{42 \zeta_3}{5}\right) + \alpha_2 \alpha_4^2 \left(\frac{3201}{8} + 90 \zeta_3\right) + \alpha_3 \alpha_4^2 \left(1708 - 384 \zeta_3\right)
- \alpha_4^3 \left(423 + 24 \zeta_3\right) - 120 \alpha_4^2 \alpha_7  + 144 \alpha_4 \alpha_7^2
\bigg]\,,
\end{align}
\end{widetext}
where $\zeta_3$
is the Riemann zeta function evaluated at argument~3.  It has been observed
in~\cite{Bednyakov:2015ooa,Zoller:2015tha} that at four-loop order the top
quark Yukawa corrections amount to 7\% of the corrections to
$\beta_3$. It is interesting to note that the remaining terms, computed in
this letter, cancel much of this contribution such that at the scale
$\mu=M_Z$, about 99\% of the four-loop coefficient is provided by the pure QCD
contribution. At three loops this is not the case; here the remaining terms
cancel about 40\% of the pure QCD contribution and thus have a significant
effect on the value. For this reason, the complete four-loop contribution to
$\beta_3$ provides a large correction compared to the three-loop
contributions. We find that the four-loop contributions to
$\beta_1$, $\beta_2$ and
$\beta_3$ amount to 8\%, 5\% and 127\% of the three-loop contributions.

\smallskip

{\bf Summary.}  
We compute analytic expressions for the four-loop gauge coupling beta functions in the SM,
which require a consistent treatment of $\gamma_5$ in $D=4-2\epsilon$
space-time dimensions. We circumvent this problem by exploiting the findings
of Refs.~\cite{Poole:2019txl,Poole:2019kcm}, which fix the relevant terms through
relations with known, unambiguous, lower-order results.  
Our calculation neglects the Yukawa
contributions from the first and second generations, which are numerically
small; their inclusion would not pose any practical problem.
The calculation performed in this letter represents the highest full-SM loop
calculation of phenomenologically relevant quantities to date.


\smallskip

{\bf Acknowledgements.} This research was supported by the Deutsche
Forschungsgemeinschaft (DFG, German Research Foundation) under grant 396021762
--- TRR 257 ``Particle Physics Phenomenology after the Higgs Discovery'' and
AET is partially supported by the Danish National Research Foundation under
grant DNRF:90.  FH acknowledges the support of the DFG-funded Doctoral School
KSETA.  CP thanks $\text{CP}^{3}$-Origins, where most of his contributions to
this project were carried out under grant DNRF:90.


\end{document}